\documentclass[reprint,superscriptaddress,amsmath,amssymb,aps,prl]{revtex4-2}

\usepackage{graphicx}
\usepackage[version=4]{mhchem}
\usepackage[dvipsnames]{xcolor}
\usepackage{soul}

\definecolor{dartmouthgreen}{rgb}{0.05, 0.5, 0.06}

\begin{document}

\title{Optimization of ionic configurations in battery materials by quantum annealing}

\author{Tobias Binninger}
\email{t.binninger@fz-juelich.de}
\affiliation{Theory and Computation of Energy Materials (IET-3), Institute of Energy Technologies, Forschungszentrum J\"ulich GmbH, 52425 J\"ulich, Germany}
\affiliation{J\"ulich Aachen Research Alliance JARA Energy
\& Center for Simulation and Data Science (CSD), 52425 J\"ulich, Germany}

\author{Yin-Ying Ting}
\affiliation{Theory and Computation of Energy Materials (IET-3), Institute of Energy Technologies, Forschungszentrum J\"ulich GmbH, 52425 J\"ulich, Germany}
\affiliation{J\"ulich Aachen Research Alliance JARA Energy
\& Center for Simulation and Data Science (CSD), 52425 J\"ulich, Germany}

\author{Piotr M. Kowalski}
\affiliation{Theory and Computation of Energy Materials (IET-3), Institute of Energy Technologies, Forschungszentrum J\"ulich GmbH, 52425 J\"ulich, Germany}
\affiliation{J\"ulich Aachen Research Alliance JARA Energy
\& Center for Simulation and Data Science (CSD), 52425 J\"ulich, Germany}

\author{Michael H. Eikerling}
\affiliation{Theory and Computation of Energy Materials (IET-3), Institute of Energy Technologies, Forschungszentrum J\"ulich GmbH, 52425 J\"ulich, Germany}
\affiliation{J\"ulich Aachen Research Alliance JARA Energy
\& Center for Simulation and Data Science (CSD), 52425 J\"ulich, Germany}
\affiliation{Chair of Theory and Computation of Energy Materials, Faculty of Georesources and Materials Engineering, RWTH Aachen University, Intzestrasse 5, 52072 Aachen, Germany}

\begin{abstract}
Energy materials with disorder in site occupation are challenging for computational studies due to an exponential scaling of the configuration space. We herein present a grand-canonical optimization method that enables the use of quantum annealing (QA) for sampling the ionic ground state. The method relies on a Legendre transformation of the Coulomb energy cost function that strongly reduces the effective coupling strengths of the fully connected problem, which is essential for effectiveness of QA. The approach is expected to be applicable to a variety of materials optimization problems.
\end{abstract}

\maketitle

Modeling of ionic arrangements in multi-elements compounds represents a ubiquitous challenge for computational research in energy materials. Materials with mixed or partially occupied lattice sites are widely investigated, e.g., doped semiconductors for photovoltaics~\cite{chenCompositionalDependenceStructural2011,guTinMixedLead2020,helmers68EfficientGaAsBased2021}, or intercalation materials and ionic conductors for Li-ion batteries (LIBs)~\cite{ohzukuLayeredLithiumInsertion2001, sarkarHighEntropyOxides2018, muruganFastLithiumIon2007, berardanRoomTemperatureLithium2016}. While the configurational arrangement of elements impacts computed thermodynamic~\cite{liuSpecialQuasiorderedStructures2016,FKK17}, electronic~\cite{yangBandStructureEngineering2018}, chemical~\cite{chaeEffectsLocalCompositional2022}, and ionic-transport parameters \cite{TCB21,BVK21}, construction of reliable models of occupation disorder represents a major difficulty for simulations~\cite{sanchezGeneralizedClusterDescription1984,laksEfficientClusterExpansion1992,zungerSpecialQuasirandomStructures1990,vandewalleAlloyTheoreticAutomated2002,lerchUNCLECodeConstructing2009,okhotnikovSupercellProgramCombinatorial2016}. 

For a simulation cell comprising $M$ sites, a fraction $\theta$ of which being occupied, the total number of possible configurations is given by (using Stirling's formula)
\begin{equation} 
\binom{M}{\theta M} \approx \left[\theta^{-\theta}(1-\theta)^{-(1-\theta)}\right]^M \ . 
\end{equation} 
The exponential scaling of the configuration space with system size ($M$) precludes an efficient sampling of all possible configurations. For many computational problems, thermodynamically relevant low(est)-energy configurations must be computed to reliably predict materials properties. Finding such ionic distributions requires efficient algorithms. Common approaches to this end include the methods of cluster expansion~\cite{sanchezGeneralizedClusterDescription1984, laksEfficientClusterExpansion1992, vandewalleAlloyTheoreticAutomated2002, lerchUNCLECodeConstructing2009, huangFindingProvingExact2016} and special quasirandom structures~\cite{zungerSpecialQuasirandomStructures1990, okhotnikovSupercellProgramCombinatorial2016}. Alternatively, computational workflows employ stochastic Monte Carlo methods to identify a number of candidate ground-state configurations~\cite{mottetDopingGarnettypeElectrolytes2019, binningerComparisonComputationalMethods2020}. The obtained models are used for accurate and computationally intensive simulations, usually based on density functional theory (DFT) (e.g., \cite{binningerComparisonComputationalMethods2020}). For typical simulation cells of LIB materials, containing less than a hundred intercalation sites, the total number of ionic configurations is of the order of $10^{9}$--$10^{15}$ and the sampling of ionic configurations significantly contributes to the runtime of computational workflows. Efficient sampling methods are thus needed to simulate the charging/discharging characteristics of LIB cathodes.

Quantum computing (QC) techniques provide new ways of solving exponentially scaling problems in materials science. Among these, quantum annealing (QA), a type of adiabatic quantum computing~\cite{albashAdiabaticQuantumComputation2018}, is designed to solve classical optimization problems, which can be mapped onto an Ising-type Hamiltonian~\cite{haukePerspectivesQuantumAnnealing2020, jungerQuantumAnnealingDigital2021, symonsPractitionersGuideQuantum2023}. The underlying procedure consists of adiabatically tuning an initial transverse-field Hamiltonian to the target Hamiltonian, encoding the cost function of the optimization problem. Then, the quantum state of the multi-qubit system adiabatically converges to the ground state of the target Hamiltonian, corresponding to the global minimum of the cost function. With the emergence of commercially available hardware, such as the QA devices by D-Wave Systems Inc., such methods should be tested and deployed in materials science research. The efficiency of QA depends on finding a suitable encoding of a given optimization problem on the QA hardware~\cite{chancellorDomainWallEncoding2019, 2021_IEEE_Chancellor}. Due to limited connectivity of the qubit network topology, this is particularly challenging for problems represented by a fully connected interaction graph~\cite{pelofskeSolvingLargerMaximum2023}.

QA and QA-inspired approaches have been applied for conformational sampling of polymer mixtures~\cite{michelettiPolymerPhysicsQuantum2021}, crystal-structure prediction~\cite{gusevOptimalityGuaranteesCrystal2023}, and materials design and optimization~\cite{kitaiDesigningMetamaterialsQuantum2020, hatakeyama-satoTacklingChallengeHuge2021, choubisaAcceleratedChemicalSpace2023}. Carnevali et al.~\cite{carnevaliVacanciesGrapheneApplication2020} and Camino et al.~\cite{caminoQuantumComputingMaterials2023} employed D-Wave QA devices for optimizing the distribution of vacancies in a graphene sheet. In their approach, each site $i$ of the graphene lattice was represented by a binary site occupation variable $x_i\in \{0,1\}$, indicating whether respective site is occupied or vacant. The energy cost function accounted for the number of intact vs. broken chemical bonds. The limited number of chemical bonds per atom resulted in a limited number of non-zero $x_i x_j$ coupling terms, which was beneficial for the mapping to the limited connectivity of the D-Wave qubit network. Optimization of the bare energy model resulted in the complete occupation of all sites. To tune the system to a certain number of vacancies, a term $\left(\sum_i x_i - N_C\right)^2$ was added to the cost function, penalizing states that deviate from the targeted atom number, $N_C$. However, this penalty term produces non-zero couplings of all pairs of variables, thus thwarting the sparse form of the bond energy model.

Herein, we demonstrate the use of quantum annealing for sampling the configurational ground state of ionic materials, employing the total Coulomb energy as a surrogate energy model~\cite{mottetDopingGarnettypeElectrolytes2019, binningerComparisonComputationalMethods2020}. The long-range nature of Coulomb interactions couples any pair of lattice sites. The resulting full connectivity of the optimization problem is further exacerbated by the penalty term for the target stoichiometry constraint, making the problem extremely challenging for present-day QA architectures. To overcome this difficulty, we propose a grand-canonical optimization method employing a Legendre-transformed energy cost function that significantly alleviates the connectivity strength. The method renders the configurational optimization of LIB materials feasible on existing D-Wave QA hardware and has general applicability to similar problems in the research of disordered materials. While we do not claim the QA method to outperform classical algorithms at present stage, pronounced performance gains of quantum computing methods are reasonable to expect in the future.

\begin{figure}[t]
\centering
\includegraphics[width=\columnwidth]{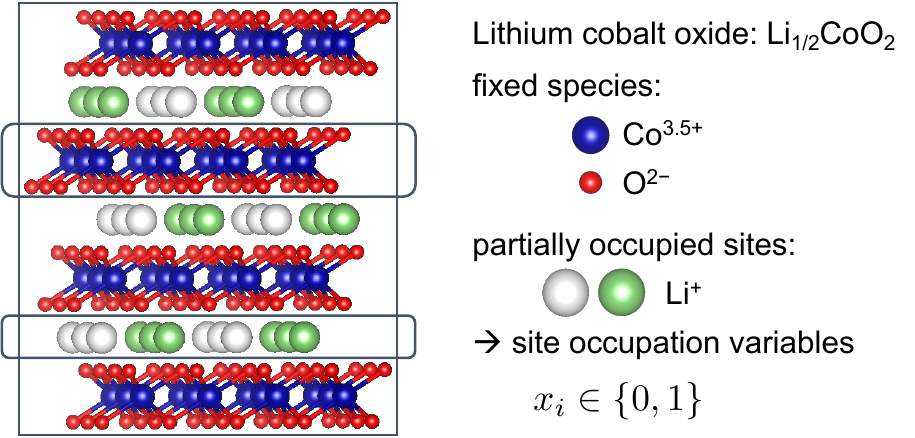}
\caption{Layered crystal structure of lithium cobalt oxide (LCO). The ground-state configuration of the Coulomb energy model is shown, with white and green spheres representing vacant and occupied Li sites, respectively.}
\label{fig_1}
\end{figure}

We have chosen lithium cobalt oxide (LCO), a standard cathode material for LIBs~\cite{mizushimaLixCoO21980}, as a test case. During charging/discharging of LCO, Li ions are extracted/intercalated according to the reaction: 
\begin{equation} 
\ce{LiCoO2} \ \rightleftarrows\ \ce{Li_{1-y}CoO2} + y\,\ce{Li^+} + y\,\ce{e^-} \ , 
\end{equation}
resulting in the formation/filling of vacancies across the Li-ion sub-lattice. We target to model the semi-lithiated state with \ce{Li_{0.5}CoO2} stoichiometry. The model cell, shown in Fig.~\ref{fig_1}, comprises 36 Li sites, half of which are occupied and half are vacant. The problem consists in finding the ground-state configuration among the $\approx 10^{10}$ possible distributions of 18 Li ions over 36 available sites. This task is sufficiently complex for assessing the QA performance while being amenable for benchmarking against a classical method, such as Replica Exchange Monte Carlo (REMC).

The electrostatic Coulomb energy, 
\begin{equation} 
E_{\mathrm{coul}} = \frac{e^2}{4\pi\epsilon_0}\,\sum_{\alpha<\beta} \frac{Z_{\alpha} Z_{\beta}}{|r_{\alpha}-r_{\beta}|} \ ,
\label{coul}
\end{equation} 
is the cost function to be minimized. Here, the summation goes over any pair of ions present in the lattice, $Z_{\alpha}$ are the respective valencies, and other constants have their usual meaning. Standard valencies of $Z_{\ce{Li}} = +1$ and $Z_{\ce{O}} = -2$ have been chosen for lithium cations and oxygen anions, respectively, and $Z_{\ce{Co}} = +3.5$, which is the (average) valency of cobalt cations in semi-lithiated LCO to provide overall charge neutrality. The Coulomb model is a crude approximation of the system's energy, especially with respect to electronic contributions. In reality, LCO reveals a complex interplay between (de)lithiation and electronic structure~\cite{marianettiFirstorderMottTransition2004}. For the purpose of the present work, the Coulomb energy serves only as a convenient surrogate model to study ionic configurational optimization. 

Assigning binary occupation variables $x_i$ to each of the Li sites, indicating whether a given site is occupied ($x_i = 1$) or vacant ($x_i = 0$), the Coulomb energy can be written in terms of sums over all Li sites,
\begin{align}
E_{\mathrm{coul}} & = \mathrm{const} + \sum_{i\,\in\,\mathcal{S}_{\mathrm{Li}}} Q_{i,i}\,x_i + \sum_{i<j\,\in\,\mathcal{S}_{\mathrm{Li}}} Q_{i,j}\,x_i\,x_j \ .
\label{eq_Coulomb_QUBO}
\end{align}
Here, $\mathrm{const} = \frac{e^2}{4\pi\epsilon_0}\,\sum_{i<j\,\in\,\mathrm{fix}} \frac{Z_i Z_j}{|r_i-r_j|}$ is the Coulomb interaction energy among all fixed ions, i.e., cobalt cations and oxygen anions, and the coefficients $Q_{i,i} = \frac{e^2}{4\pi\epsilon_0}\,\sum_{j\,\in\,\mathrm{fix}} \frac{Z_j}{|r_i-r_j|}$ and $Q_{i,j} = \frac{e^2}{4\pi\epsilon_0}\,\frac{1}{|r_i-r_j|}$ correspond to the Coulomb interaction between a given Li site $i$ and fixed species, and between a given pair of Li sites, respectively. Due to the pairwise character of Coulomb interactions, Eq.~\eqref{eq_Coulomb_QUBO} has the form of a quadratic unconstrained binary optimization (QUBO) problem, as required for present D-Wave QA devices.

\begin{figure*}[t]
\centering
\includegraphics[width=2\columnwidth]{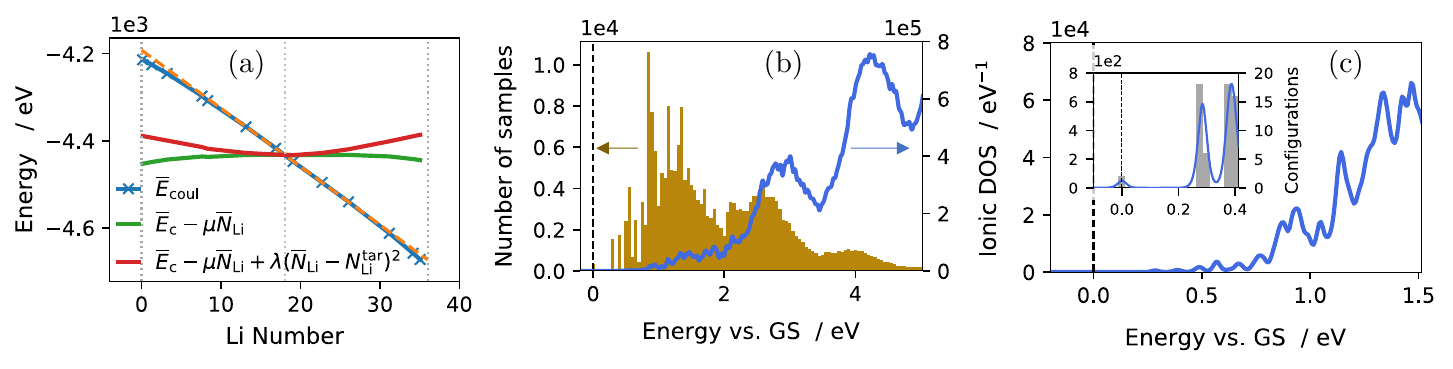}
\caption{(a) Average energy vs. Li number for LCO as obtained from the bare Coulomb energy model of Eq.~\eqref{eq_Coulomb_QUBO} (blue curve with markers), after applying a Legendre transformation with $\mu = -13.38\,\mathrm{eV}$ (green curve), and including the quadratic penalty term of Eq.~\eqref{eq_penalty_stoichio_constraint} with $\lambda = 0.2$ (red curve). (b) Histogram of Coulomb energies obtained from QA with the grand-canonical method ($\mu = -13.2\,\mathrm{eV}$ and $\lambda = 0.2$). Only samples with $N_{\mathrm{Li}} = 18$ were counted. The energy is given vs. the ground-state (GS) energy ($E_{\mathrm{coul}}^{\mathrm{min}} = -4432.64\,\mathrm{eV}$), i.e., what is shown is the difference $E_{\mathrm{coul}} - E_{\mathrm{coul}}^{\mathrm{min}}$. The ionic configurational density of states (DOS) of the Coulomb energy model, determined by REMC sampling, is shown for comparison (blue curve). (c) Ionic DOS, obtained from REMC sampling, in the energy range around the GS (blue: DOS curve with thermal broadening, left-hand axis; grey bars in inset: DOS histogram, right-hand axis).}
\label{fig_2}
\end{figure*}

Due to periodic boundary conditions (PBC), each variable $x_i$ represents a given Li site of the model cell plus all of its periodic images. To ensure that long-range Coulomb interactions are properly accounted for, we employed Ewald summation routines available in the \texttt{pymatgen} library~\cite{ongPythonMaterialsGenomics2013} for \texttt{Python} in computing the QUBO coefficients $Q_{i,i}$ and $Q_{i,j}$ under PBC. The constant term was obtained as the Ewald energy of the simulation cell with only fixed species present. To determine $Q_{i,i}$, simulation cells with only one occupied Li site $i$ and all fixed ion species were constructed. The respective Ewald energies were corrected by subtracting the constant term to avoid over-counting of the interaction energy among fixed species. The coefficients $Q_{i,j}$ were obtained from the Ewald energies of simulation cells with only Li ions on sites $i$ and $j$ present (without any fixed ions). To avoid double counting, the respective energies were corrected for the self-energies of sites $i$ and $j$, i.e., the interaction energy of Li ions on a single given site and all of its periodic images (which is already accounted for in the respective diagonal terms $Q_{i,i}$). We note that simulation cells comprising only a subset of ion species are not charge balanced. The Ewald method automatically adds a neutralizing homogeneous background charge to prevent divergence of the electrostatic energy. However, due to charge neutrality of the overall system, background charge contributions mutually get cancelled and do not affect the total Coulomb energy of Eq.~\eqref{eq_Coulomb_QUBO}. The obtained QUBO coefficients for LCO were: $\mathrm{const} = -4212.68\,\mathrm{eV}$, $Q_{i,i} = -9.40\,\mathrm{eV}$, and 
$-1.04\,\mathrm{eV} \leq Q_{i,j} \leq 2.02\,\mathrm{eV}$ for $i<j$ ($Q_{i,j} = 0$ for $i>j$).

\begin{table}
\caption{\label{tab_param_tuning} Parameter tuning for sampling the ionic ground-state configuration on a D-Wave Advantage\texttrademark{} system. For each set of parameters, 1000 independent annealing runs were performed with an annealing time of $100\,\mathrm{\mu s}$ each. $\lambda$: strength of Li number constraint; $\mu$: chemical potential; $\bar{N}_{\ce{Li}}$: average Li number; $\sigma_{\mathrm{c}}$: chain strength; $\eta_{\mathrm{c}}$: fraction of broken chains; $E_{\mathrm{coul}}^{\mathrm{min}}$: minimum value of Coulomb energy for configurations with target stoichiometry.}
\begin{ruledtabular}
\begin{tabular}{cccccc}
  $\lambda$ & $\mu$ & $\bar{N}_{\ce{Li}}$ & $\sigma_{\mathrm{c}}$ & $\eta_{\mathrm{c}}$ & $E_{\mathrm{coul}}^{\mathrm{min}}$ \\[0.1cm]
  \hline
   0 & -- & 36 & 1 & 0.02\% & -- \\
   0.5 & -- & 34.9 & 1 & 49\% & -- \\
   1.0 & -- & 23.8 & 1 & 92\% & -- \\
   1.0 & -- & 24.3 & 5 & 0.03\% & -- \\
   5.0 & -- & 19.2 & 50 & 0.02\% & $-4427.08\,\mathrm{eV}$ \\
   0.2 & $-13.2\,\mathrm{eV}$ & 18.0 & 2 & 0.03\% & $-4432.64\,\mathrm{eV}$ \\
\end{tabular}
\end{ruledtabular}
\end{table}

We performed optimization of the obtained QUBO function using the D-Wave Advantage\texttrademark{} QA system. Due to pairwise Coulomb interactions, the problem is fully connected, and we thus employed the \texttt{DWaveCliqueSampler()} routine from the D-Wave Ocean library. Minimization of the bare Coulomb energy cost function resulted in $x_i = 1$ for all 36 Li sites, i.e., complete occupation of the Li sublattice. Unlike in classical sampling algorithms, the search space cannot be restricted to the target stoichiometry in QA. To obtain the semi-lithiated state with desired occupation of $N_{\mathrm{Li}}^{\mathrm{target}} = 18$ sites, the cost function must be modified to penalize configurations that violate the target stoichiometry. The standard approach to enforce such constraint consists in adding a penalty term~\cite{carnevaliVacanciesGrapheneApplication2020, caminoQuantumComputingMaterials2023}: 
\begin{align}
\lambda \left( \sum_{i} x_i - N_{\mathrm{Li}}^{\mathrm{target}} \right)^2 \ ,
\label{eq_penalty_stoichio_constraint}
\end{align}
where $\sum_{i} x_i = N_{\mathrm{Li}}$, also known as the Hamming weight of the binary string, is the total number of Li ions for a given configuration and $\lambda$ is a parameter controlling the strength of the constraint. Sampling statistics obtained for different values of $\lambda$ are shown in Table~\ref{tab_param_tuning}. With increasing $\lambda$, the average Li number $\bar{N}_{\ce{Li}}$ of the output configurations of 1000 independent annealing runs decreased towards the desired value of 18, but at the same time the fraction of broken qubit chains~\footnote{The D-Wave Advantage\texttrademark{} system employs the so-called Pegasus network topology with $\approx 15$ connections per qubit. Depending on the connectivity of the QUBO problem, each logical variable must be represented by a group, or chain, of physical qubits. All qubits of a given chain must return the same value, which is achieved by tuning the respective coupling strengths via the chain strength parameter $\sigma_{\mathrm{c}}$.}, $\eta_{\mathrm{c}}$, increased to 92\% for $\lambda = 1$, rendering the solutions unreliable. This could be prevented by concomitantly increasing the chain strength parameter $\sigma_{\mathrm{c}}$. For $\lambda = 5$ and $\sigma_{\mathrm{c}} = 50$, we have achieved a negligible fraction of chain breaks and obtained the target stoichiometry of $N_{\mathrm{Li}}^{\mathrm{target}} = 18$ in 16\% of annealing runs. However, the respective minimum value of the Coulomb energy of $-4427.08\,\mathrm{eV}$ is significantly larger than the minimum energy of $-4432.64\,\mathrm{eV}$ obtained with classical REMC sampling. The reason for the poor performance of the QA method lies in the large value of $\lambda = 5$ required for the stoichiometry constraint, which adds a value of $2\lambda = 10$ to the off-diagonal elements of the coefficient matrix, a factor of 5--10 larger in magnitude than the off-diagonal contributions resulting from the Coulomb energy. The stoichiometry constraint thus effectively masks the Coulomb energy terms and renders the optimization inefficient.

To meet the Li target stoichiometry at much smaller bias of the off-diagonal elements, we introduce a grand-canonical optimization method. Fig.~\ref{fig_2}a shows the average energy of sampled configurations as a function of the average Li number (blue curve with markers)~\footnote{The data points plotted in Fig.~\ref{fig_2}a were obtained from sampling runs using the cost function of Eq.~\eqref{eq_grand_canonical_cost} with $\lambda = 0$ and different values of $\mu \in [-20\,\mathrm{eV},-10\,\mathrm{eV}]$. For each value of $\mu$, the average Coulomb energy of the sampling output was plotted vs. the corresponding average Li number.}. A negative slope is apparent, explaining (i) why minimization of the bare energy resulted in complete lithiation and (ii) why a large quadratic penalty $\lambda$ is required to enforce a minimum of the cost function close to the target stoichiometry of 18. Within the context of the Coulomb energy model, we interpret the local slope at the target stoichiometry as the chemical potential, $\mu = \partial \bar{E}_{\mathrm{coul}}/\partial\bar{N}_{\ce{Li}}$, and rotate the energy curve by performing a Legendre transformation from Coulomb energy to the grand-canonical cost function $E_{\mathrm{coul}}-\mu N_{\mathrm{Li}}$. Using the fitted value of $\mu = -13.38\,\mathrm{eV}$ (dashed line in Fig.~\ref{fig_2}a), the cost function becomes flat around the target stoichiometry with a slightly negative curvature (green curve in Fig.~\ref{fig_2}a)~\footnote{We consider the slightly negative curvature to result from the compensating background charge implicitly applied for systems with $N_{\mathrm{Li}} \neq 18$. In a real system, charge compensation is provided by a reduction/oxidation of the active transition metal ion species and the respective chemical hardness introduces a positive energy curvature~\cite{saubanereIntuitiveEfficientMethod2014}. Moreover, while Fig.~\ref{fig_2}a shows the average energy, the \emph{free energy} contains an additional entropic contribution producing an overall positive curvature as required for thermodynamic stability.}. Then, the quadratic penalty term of Eq.~\eqref{eq_penalty_stoichio_constraint} with a small value of $\lambda = 0.2$ is sufficient to bend the cost function upwards and produce a minimum at the target Li number (red curve in Fig.~\ref{fig_2}a). We note that the grand-canonical transformation only depends on the total Li number and therefore does not interfere with the energy optimization at the target stoichiometry. Moreover, being a linear transformation, it only shifts the diagonal elements of the QUBO matrix by a constant $-\mu$, thus avoiding the aforementioned problem of masking of off-diagonal elements. The total cost function for grand-canonical optimization thus reads
\begin{align}
E_{\mathrm{coul}}[\{x_i\}] - \mu \sum_{i\,\in\,\mathcal{S}_{\mathrm{Li}}} x_i + \lambda \left( \sum_{i\,\in\,\mathcal{S}_{\mathrm{Li}}} x_i - N_{\mathrm{Li}}^{\mathrm{target}} \right)^2  ,
\label{eq_grand_canonical_cost}
\end{align}
where $E_{\mathrm{coul}}[\{x_i\}]$ is given by Eq.~\eqref{eq_Coulomb_QUBO}. Here, the objective still consists in identifying the energetic \emph{ground state} at a given target stoichiometry, not to be confused with methods for \emph{thermodynamic} sampling~\cite{1998_VanderVen_PhysRevB}.

Applying this method with a fine-tuned chemical potential of $\mu = -13.2\,\mathrm{eV}$, we obtained significantly better performance of the QA procedure, cf. Table~\ref{tab_param_tuning}. The target stoichiometry of $N_{\mathrm{Li}} = 18$ was met in 55\% of returned configurations at a mild value of $\lambda = 0.2$. Most importantly, the minimum of returned Coulomb energies at the target stoichiometry was $E_{\mathrm{coul}}^{\mathrm{min}} = -4432.64\,\mathrm{eV}$, which is identical to the minimum energy obtained from the benchmark REMC sampling. The proposed grand-canonical method thus makes the fully connected Coulomb energy model feasible for optimization by QA, which is the main result of the present work. The respective ground-state configuration is shown in Fig.~\ref{fig_1}. The row-like Li ordering is in agreement with previous computational and experimental findings for semi-lithiated LCO~\cite{1998_VanderVen_PhysRevB, wolvertonFirstPrinciplesPredictionVacancy1998,  1992_JECS_Reimers}.

\begin{figure*}[t]
\centering
\includegraphics[width=2\columnwidth]{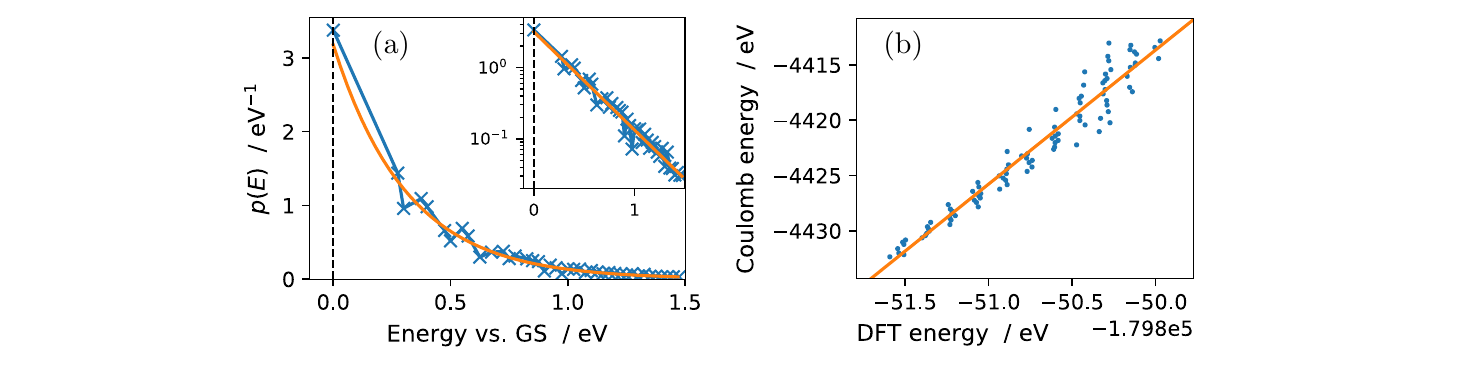}
\caption{(a) Intrinsic sampling probability (per energy), $p(E)$, obtained by dividing the overall sampling rate by the configurational DOS. Orange fitted curve: Boltzmann-type exponential $\exp(-E/kT)$. Inset: Same on a logarithmic scale. (b) Comparison of the Coulomb energy vs. DFT energy for 100 different ionic configurations of semi-lithiated LCO. DFT method: Calculations performed with Quantum Espresso software package~\cite{giannozziQUANTUMESPRESSOModular2009}; Ultrasoft pseudopotentials~\cite{vanderbiltSoftSelfconsistentPseudopotentials1990} with GGA-PBEsol~\cite{perdewRestoringDensityGradientExpansion2008} exchange-correlation functional; DFT+$U$ method with Hubbard parameter $U=4.6\,\mathrm{eV}$ for cobalt~\cite{tingRefinedDFTMethod2023}; Cutoff energy of 50\,Ry for plane-wave basis set; $3 \times 4 \times 2$ $k$-point mesh.}
\label{fig_3}
\end{figure*}

We finally performed a deeper analysis of the D-Wave sampling statistics. Fig.~\ref{fig_2}b presents a histogram of the Coulomb energies returned from 400'000 annealing runs (only using results with $N_{\mathrm{Li}} = 18$
and without any chain breaks). A broad distribution of energies is obtained, with most of the samples being few eV above the ground state (GS), whereas the true minimum energy solution was returned in only 0.083\% of annealing runs. At first glance, this appears to be a rather low optimization efficiency. However, the configurational density of states (DOS) of the underlying model must be analysed for a fair assessment of the statistics~\cite{bruggerOutputStatisticsQuantum2022}. The ionic DOS of the Coulomb energy model, $N_{\mathrm{DOS}}(E)$, was obtained from extended REMC sampling runs. It is shown as a blue curve in Fig.~\ref{fig_2}b, with close-ups around the GS energy in Fig.~\ref{fig_2}c. The essential structure of the QA sampling histogram (golden bars in Fig.~\ref{fig_2}b) reflects the shape of the underlying DOS. Assuming that each configuration is sampled with a certain ``intrinsic'' probability, $p(E)$, that only depends on the respective energy, the overall sampling rate, $N(E)$, is proportional to $p(E)$ times the number of states with energy $E$, i.e., the configurational DOS,
\begin{align}
N(E) \,\propto\, p(E)\,N_{\mathrm{DOS}}(E) \ .
\label{eq_sampling_prob}
\end{align}
To extract $p(E)$, we normalized the QA sampling histogram with the configurational DOS. The result is shown in Fig.~\ref{fig_3}a. A monotonically decreasing probability as a function of energy is obtained, clearly indicating that the lower energy configurations are sampled with higher probability, with the ground-state configuration having the highest sampling probability. The obtained $p(E)$ curve is well reproduced by a Boltzmann-type exponential, $\exp(-E/kT)$, with a fitted value of $kT = 0.31\,\mathrm{eV}$ (orange curve). Such QA statistics has been observed previously~\cite{benedettiEstimationEffectiveTemperatures2016, benedettiQuantumAssistedLearningHardwareEmbedded2017, bruggerOutputStatisticsQuantum2022} and explained by statistical imperfections in tuning the target Hamiltonian~\cite{bruggerOutputStatisticsQuantum2022}. We note that the \emph{effective} sampling temperature depends on the problem at hand and is not related to the physical temperature of the hardware~\cite{benedettiEstimationEffectiveTemperatures2016}. 

At first glance, our fitted value of $kT = 0.31\,\mathrm{eV}$ indicates ``hot'' sampling of the configurational space. However, the effective temperature scales with the energy scale of the problem. The Coulomb energy of Eq.~\eqref{eq_Coulomb_QUBO} represents a hard energy model, because it neglects dielectric screening. Including the latter in the form of a dielectric constant, $\epsilon_r$, scales down all energies, and thus the effective sampling temperature. To estimate $\epsilon_r$, we have computed the DFT energies of 100 randomly selected ionic configurations. Fig.~\ref{fig_3}b reveals a linear correlation between the Coulomb and corresponding DFT energies, which demonstrates the physical meaningfulness of the ionic Coulomb energy model for LCO. Since DFT energies implicitly include the effect of electronic screening, we interpret the slope of the plot in Fig.~\ref{fig_3}b as an effective dielectric constant, $\epsilon_r = 12$, which reduces the effective sampling temperature to $kT/\epsilon_r = 0.026\,\mathrm{eV}$, i.e., room-temperature.

In summary, we have presented here an efficient grand-canonical optimization method, which renders quantum annealing feasible for sampling the ionic ground state based on a fully interacting Coulomb energy model. The method has been demonstrated on a D-Wave Advantage\texttrademark{} quantum annealer to successfully identify the lowest energy arrangement of lithium ions in lithium cobalt oxide. Boltzmann-type output statistics was observed with the highest sampling probability for the ground state configuration. We consider the grand-canonical optimization method to be of more general applicability to the solution of materials optimization problems by quantum computing. \\

The authors gratefully acknowledge the J\"ulich Supercomputing Centre (\url{https://www.fz-juelich.de/ias/jsc}) for funding this project by providing computing time on the D-Wave Advantage\texttrademark{} System JUPSI through the J\"ulich UNified Infrastructure for Quantum computing (JUNIQ) within the project qdisk. DFT simulations were performed on the JURECA machine in the scope of the project cjiek61. The presented work was carried out within the framework of the Helmholtz Association’s program Materials and Technologies for the Energy Transition, Topic 2: Electrochemical Energy Storage.


\begin{thebibliography}{10}

\bibitem{chenCompositionalDependenceStructural2011}
Shiyou Chen, Aron Walsh, Ji-Hui Yang, X.~G. Gong, Lin Sun, Ping-Xiong Yang,
  Jun-Hao Chu, and Su-Huai Wei.
\newblock Compositional dependence of structural and electronic properties of
  \ce{Cu2ZnSn(S,Se)4} alloys for thin film solar cells.
\newblock {\em Phys. Rev. B}, 83(12):125201, 2011.

\bibitem{guTinMixedLead2020}
Shuai Gu, Renxing Lin, Qiaolei Han, Yuan Gao, Hairen Tan, and Jia Zhu.
\newblock Tin and {{Mixed Lead}}\textendash{{Tin Halide Perovskite Solar
  Cells}}: {{Progress}} and their {{Application}} in {{Tandem Solar Cells}}.
\newblock {\em Advanced Materials}, 32(27):1907392, 2020.

\bibitem{helmers68EfficientGaAsBased2021}
Henning Helmers, Esther Lopez, Oliver H{\"o}hn, David Lackner, Jonas Sch{\"o}n,
  Meike Schauerte, Michael Schachtner, Frank Dimroth, and Andreas~W. Bett.
\newblock 68.9\% {{Efficient GaAs-Based Photonic Power Conversion Enabled}} by
  {{Photon Recycling}} and {{Optical Resonance}}.
\newblock {\em physica status solidi (RRL) \textendash{} Rapid Research
  Letters}, 15(7):2100113, 2021.

\bibitem{ohzukuLayeredLithiumInsertion2001}
Tsutomu Ohzuku and Yoshinari Makimura.
\newblock Layered {{Lithium Insertion Material}} of
  \ce{LiCo_{1/3}Ni_{1/3}Mn_{1/3}O2} for {{Lithium-Ion Batteries}}.
\newblock {\em Chem. Lett.}, 30(7):642--643, 2001.

\bibitem{sarkarHighEntropyOxides2018}
Abhishek Sarkar, Leonardo Velasco, Di~Wang, Qingsong Wang, Gopichand Talasila,
  Lea {de Biasi}, Christian K{\"u}bel, Torsten Brezesinski, Subramshu~S.
  Bhattacharya, Horst Hahn, and Ben Breitung.
\newblock High entropy oxides for reversible energy storage.
\newblock {\em Nat Commun}, 9(1):3400, 2018.

\bibitem{muruganFastLithiumIon2007}
Ramaswamy Murugan, Venkataraman Thangadurai, and Werner Weppner.
\newblock Fast {{Lithium Ion Conduction}} in {{Garnet-Type \ce{Li7La3Zr2O12}}}.
\newblock {\em Angewandte Chemie International Edition}, 46(41):7778--7781,
  2007.

\bibitem{berardanRoomTemperatureLithium2016}
D.~B{\'e}rardan, S.~Franger, A.~K. Meena, and N.~Dragoe.
\newblock Room temperature lithium superionic conductivity in high entropy
  oxides.
\newblock {\em J. Mater. Chem. A}, 4(24):9536--9541, 2016.

\bibitem{liuSpecialQuasiorderedStructures2016}
Jian Liu, Maria~V. {Fern{\'a}ndez-Serra}, and Philip~B. Allen.
\newblock Special quasiordered structures: {{Role}} of short-range order in the
  semiconductor alloy \ce{(GaN)_{1-x}(ZnO)_{x}}.
\newblock {\em Phys. Rev. B}, 93(5):054207, 2016.

\bibitem{FKK17}
S.~Finkeldei, Ph. Kegler, P.M. Kowalski, C.~Schreinemachers, F.~Brandt, A.A.
  Bukaemskiy, V.L. Vinograd, G.~Beridze, A.~Shelyug, A.~Navrotsky, and
  D.~Bosbach.
\newblock Composition dependent order-disorder transition in
  $\mathrm{Nd}_\mathrm{x}\mathrm{Zr}_{1-\mathrm{x}}\mathrm{O}_{2-0.5\mathrm{x}}$
  pyrochlores: A combined structural, calorimetric and ab initio modeling
  study.
\newblock {\em Acta Mater.}, 125:166 -- 176, 2017.

\bibitem{yangBandStructureEngineering2018}
Jingxiu Yang, Peng Zhang, and Su-Huai Wei.
\newblock Band {{Structure Engineering}} of {{\ce{Cs2AgBiBr6} Perovskite}}
  through {{Order}}\textendash{{Disordered Transition}}: {{A First-Principle
  Study}}.
\newblock {\em J. Phys. Chem. Lett.}, 9(1):31--35, 2018.

\bibitem{chaeEffectsLocalCompositional2022}
Sieun Chae, Logan Williams, Jihang Lee, John~T. Heron, and Emmanouil Kioupakis.
\newblock Effects of local compositional and structural disorder on vacancy
  formation in entropy-stabilized oxides from first-principles.
\newblock {\em npj Comput Mater}, 8(1):1--7, 2022.

\bibitem{TCB21}
Timothy Connor, Oskar Cheong, Thomas Bornhake, Alison~C. Shad, Rebekka Tesch,
  Mengli Sun, Zhengda He, Andrey Bukayemsky, Victor~L. Vinograd, Sarah~C.
  Finkeldei, and Piotr~M. Kowalski.
\newblock Pyrochlore compounds from atomistic simulations.
\newblock {\em Frontiers in Chemistry}, 9:940, 2021.

\bibitem{BVK21}
Andrey~A. Bukaemskiy, Victor~L. Vinograd, and Piotr~M. Kowalski.
\newblock Ion distribution models for defect fluorite {ZrO$_2$-AO$_{1.5}$
  (A=Ln, Y)} solid solutions: I. relationship between lattice parameter and
  composition.
\newblock {\em Acta Mater.}, 202:99--111, 2021.

\bibitem{sanchezGeneralizedClusterDescription1984}
J.~M. Sanchez, F.~Ducastelle, and D.~Gratias.
\newblock Generalized cluster description of multicomponent systems.
\newblock {\em Physica A: Statistical Mechanics and its Applications},
  128(1):334--350, 1984.

\bibitem{laksEfficientClusterExpansion1992}
David~B. Laks, L.~G. Ferreira, Sverre Froyen, and Alex Zunger.
\newblock Efficient cluster expansion for substitutional systems.
\newblock {\em Phys. Rev. B}, 46(19):12587--12605, 1992.

\bibitem{zungerSpecialQuasirandomStructures1990}
Alex Zunger, S.-H. Wei, L.~G. Ferreira, and James~E. Bernard.
\newblock Special quasirandom structures.
\newblock {\em Phys. Rev. Lett.}, 65(3):353--356, 1990.

\bibitem{vandewalleAlloyTheoreticAutomated2002}
A.~{van de Walle}, M.~Asta, and G.~Ceder.
\newblock The alloy theoretic automated toolkit: {{A}} user guide.
\newblock {\em Calphad}, 26(4):539--553, 2002.

\bibitem{lerchUNCLECodeConstructing2009}
D.~Lerch, O.~Wieckhorst, G.~L.~W. Hart, R.~W. Forcade, and S.~M{\"u}ller.
\newblock {{UNCLE}}: A code for constructing cluster expansions for arbitrary
  lattices with minimal user-input.
\newblock {\em Modelling Simul. Mater. Sci. Eng.}, 17(5):055003, 2009.

\bibitem{okhotnikovSupercellProgramCombinatorial2016}
Kirill Okhotnikov, Thibault Charpentier, and Sylvian Cadars.
\newblock Supercell program: A combinatorial structure-generation approach for
  the local-level modeling of atomic substitutions and partial occupancies in
  crystals.
\newblock {\em Journal of Cheminformatics}, 8(1):17, 2016.

\bibitem{huangFindingProvingExact2016}
Wenxuan Huang, Daniil~A. Kitchaev, Stephen~T. Dacek, Ziqin Rong, Alexander
  Urban, Shan Cao, Chuan Luo, and Gerbrand Ceder.
\newblock Finding and proving the exact ground state of a generalized {{Ising}}
  model by convex optimization and {{MAX-SAT}}.
\newblock {\em Phys. Rev. B}, 94(13):134424, 2016.

\bibitem{mottetDopingGarnettypeElectrolytes2019}
Matthieu Mottet, Aris Marcolongo, Teodoro Laino, and Ivano Tavernelli.
\newblock Doping in garnet-type electrolytes: {{Kinetic}} and thermodynamic
  effects from molecular dynamics simulations.
\newblock {\em Phys. Rev. Mater.}, 3(3):035403, 2019.

\bibitem{binningerComparisonComputationalMethods2020}
Tobias Binninger, Aris Marcolongo, Matthieu Mottet, Val{\'e}ry Weber, and
  Teodoro Laino.
\newblock Comparison of computational methods for the electrochemical stability
  window of solid-state electrolyte materials.
\newblock {\em J. Mater. Chem. A}, 8(3):1347--1359, 2020.

\bibitem{albashAdiabaticQuantumComputation2018}
Tameem Albash and Daniel~A. Lidar.
\newblock Adiabatic quantum computation.
\newblock {\em Rev. Mod. Phys.}, 90(1):015002, 2018.

\bibitem{haukePerspectivesQuantumAnnealing2020}
Philipp Hauke, Helmut~G. Katzgraber, Wolfgang Lechner, Hidetoshi Nishimori, and
  William~D. Oliver.
\newblock Perspectives of quantum annealing: Methods and implementations.
\newblock {\em Rep. Prog. Phys.}, 83(5):054401, 2020.

\bibitem{jungerQuantumAnnealingDigital2021}
Michael J{\"u}nger, Elisabeth Lobe, Petra Mutzel, Gerhard Reinelt, Franz Rendl,
  Giovanni Rinaldi, and Tobias Stollenwerk.
\newblock Quantum {{Annealing}} versus {{Digital Computing}}: {{An Experimental
  Comparison}}.
\newblock {\em ACM J. Exp. Algorithmics}, 26:1.9:1--1.9:30, 2021.

\bibitem{symonsPractitionersGuideQuantum2023}
Benjamin C.~B. Symons, David Galvin, Emre Sahin, Vassil Alexandrov, and Stefano
  Mensa.
\newblock A practitioner's guide to quantum algorithms for optimisation
  problems.
\newblock {\em J. Phys. A: Math. Theor.}, 56(45):453001, 2023.

\bibitem{chancellorDomainWallEncoding2019}
Nicholas Chancellor.
\newblock Domain wall encoding of discrete variables for quantum annealing and
  {{QAOA}}.
\newblock {\em Quantum Sci. Technol.}, 4(4):045004, 2019.

\bibitem{2021_IEEE_Chancellor}
Jie Chen, Tobias Stollenwerk, and Nicholas Chancellor.
\newblock Performance of domain-wall encoding for quantum annealing.
\newblock {\em IEEE Transactions on Quantum Engineering}, 2:1--14, 2021.

\bibitem{pelofskeSolvingLargerMaximum2023}
Elijah Pelofske, Georg Hahn, and Hristo~N. Djidjev.
\newblock Solving larger maximum clique problems using parallel quantum
  annealing.
\newblock {\em Quantum Inf Process}, 22(5):219, 2023.

\bibitem{michelettiPolymerPhysicsQuantum2021}
Cristian Micheletti, Philipp Hauke, and Pietro Faccioli.
\newblock Polymer {{Physics}} by {{Quantum Computing}}.
\newblock {\em Phys. Rev. Lett.}, 127(8):080501, 2021.

\bibitem{gusevOptimalityGuaranteesCrystal2023}
Vladimir~V. Gusev, Duncan Adamson, Argyrios Deligkas, Dmytro Antypov,
  Christopher~M. Collins, Piotr Krysta, Igor Potapov, George~R. Darling,
  Matthew~S. Dyer, Paul Spirakis, and Matthew~J. Rosseinsky.
\newblock Optimality guarantees for crystal structure prediction.
\newblock {\em Nature}, 619(7968):68--72, 2023.

\bibitem{kitaiDesigningMetamaterialsQuantum2020}
Koki Kitai, Jiang Guo, Shenghong Ju, Shu Tanaka, Koji Tsuda, Junichiro Shiomi,
  and Ryo Tamura.
\newblock Designing metamaterials with quantum annealing and factorization
  machines.
\newblock {\em Phys. Rev. Res.}, 2(1):013319, 2020.

\bibitem{hatakeyama-satoTacklingChallengeHuge2021}
Kan {Hatakeyama-Sato}, Takahiro Kashikawa, Koichi Kimura, and Kenichi Oyaizu.
\newblock Tackling the {{Challenge}} of a {{Huge Materials Science Search
  Space}} with {{Quantum-Inspired Annealing}}.
\newblock {\em Advanced Intelligent Systems}, 3(4):2000209, 2021.

\bibitem{choubisaAcceleratedChemicalSpace2023}
Hitarth Choubisa, Jehad Abed, Douglas Mendoza, Hidetoshi Matsumura, Masahiko
  Sugimura, Zhenpeng Yao, Ziyun Wang, Brandon~R. Sutherland, Al{\'a}n
  {Aspuru-Guzik}, and Edward~H. Sargent.
\newblock Accelerated chemical space search using a quantum-inspired cluster
  expansion approach.
\newblock {\em Matter}, 6(2):605--625, 2023.

\bibitem{carnevaliVacanciesGrapheneApplication2020}
Virginia Carnevali, Ilaria Siloi, Rosa Di~Felice, and Marco Fornari.
\newblock Vacancies in graphene: An application of adiabatic quantum
  optimization.
\newblock {\em Phys. Chem. Chem. Phys.}, 22(46):27332--27337, 2020.

\bibitem{caminoQuantumComputingMaterials2023}
B.~Camino, J.~Buckeridge, P.~A. Warburton, V.~Kendon, and S.~M. Woodley.
\newblock Quantum computing and materials science: {{A}} practical guide to
  applying quantum annealing to the configurational analysis of materials.
\newblock {\em Journal of Applied Physics}, 133(22):221102, 2023.

\bibitem{mizushimaLixCoO21980}
K.~Mizushima, P.~C. Jones, P.~J. Wiseman, and J.~B. Goodenough.
\newblock \ce{Li_{x}CoO2} ($0<x<-1$): {A} new cathode material for batteries of
  high energy density.
\newblock {\em Materials Research Bulletin}, 15(6):783--789, 1980.

\bibitem{marianettiFirstorderMottTransition2004}
C.~A. Marianetti, G.~Kotliar, and G.~Ceder.
\newblock A first-order {{Mott}} transition in {{LixCoO2}}.
\newblock {\em Nature Mater}, 3(9):627--631, 2004.

\bibitem{ongPythonMaterialsGenomics2013}
Shyue~Ping Ong, William~Davidson Richards, Anubhav Jain, Geoffroy Hautier,
  Michael Kocher, Shreyas Cholia, Dan Gunter, Vincent~L. Chevrier, Kristin~A.
  Persson, and Gerbrand Ceder.
\newblock Python {{Materials Genomics}} (pymatgen): {{A}} robust, open-source
  python library for materials analysis.
\newblock {\em Computational Materials Science}, 68:314--319, 2013.

\bibitem{Note1}
The D-Wave Advantage\texttrademark {} system employs the so-called Pegasus
  network topology with $\approx 15$ connections per qubit. Depending on the
  connectivity of the QUBO problem, each logical variable must be represented
  by a group, or chain, of physical qubits. All qubits of a given chain must
  return the same value, which is achieved by tuning the respective coupling
  strengths via the chain strength parameter $\sigma _{\protect \mathrm {c}}$.

\bibitem{Note2}
The data points plotted in Fig.~\ref {fig_2}a were obtained from sampling runs
  using the cost function of Eq.~\protect \eqref {eq_grand_canonical_cost} with
  $\lambda = 0$ and different values of $\mu \in [-20\protect \,\protect
  \mathrm {eV},-10\protect \,\protect \mathrm {eV}]$. For each value of $\mu $,
  the average Coulomb energy of the sampling output was plotted vs. the
  corresponding average Li number.

\bibitem{Note3}
We consider the slightly negative curvature to result from the compensating
  background charge implicitly applied for systems with $N_{\protect \mathrm
  {Li}} \protect \neq 18$. In a real system, charge compensation is provided by
  a reduction/oxidation of the active transition metal ion species and the
  respective chemical hardness introduces a positive energy curvature~\cite
  {saubanereIntuitiveEfficientMethod2014}. Moreover, while Fig.~\ref {fig_2}a
  shows the average energy, the \protect \emph {free energy} contains an
  additional entropic contribution producing an overall positive curvature as
  required for thermodynamic stability.

\bibitem{1998_VanderVen_PhysRevB}
A.~Van~der Ven, M.~K. Aydinol, G.~Ceder, G.~Kresse, and J.~Hafner.
\newblock First-principles investigation of phase stability in \ce{Li_xCoO_2}.
\newblock {\em Phys. Rev. B}, 58:2975--2987, 1998.

\bibitem{wolvertonFirstPrinciplesPredictionVacancy1998}
C.~Wolverton and Alex Zunger.
\newblock First-{{Principles Prediction}} of {{Vacancy Order-Disorder}} and
  {{Intercalation Battery Voltages}} in \ce{Li_xCoO_2}.
\newblock {\em Phys. Rev. Lett.}, 81(3):606--609, 1998.

\bibitem{1992_JECS_Reimers}
Jan~N. Reimers and J.~R. Dahn.
\newblock Electrochemical and in situ {X}‐ray diffraction studies of lithium
  intercalation in \ce{Li_xCoO_2}.
\newblock {\em Journal of The Electrochemical Society}, 139(8):2091, 1992.

\bibitem{giannozziQUANTUMESPRESSOModular2009}
Paolo Giannozzi, Stefano Baroni, Nicola Bonini, Matteo Calandra, Roberto Car,
  Carlo Cavazzoni, Davide Ceresoli, Guido~L. Chiarotti, Matteo Cococcioni,
  Ismaila Dabo, Andrea~Dal Corso, Stefano de~Gironcoli, Stefano Fabris, Guido
  Fratesi, Ralph Gebauer, Uwe Gerstmann, Christos Gougoussis, Anton Kokalj,
  Michele Lazzeri, Layla {Martin-Samos}, Nicola Marzari, Francesco Mauri,
  Riccardo Mazzarello, Stefano Paolini, Alfredo Pasquarello, Lorenzo Paulatto,
  Carlo Sbraccia, Sandro Scandolo, Gabriele Sclauzero, Ari~P. Seitsonen,
  Alexander Smogunov, Paolo Umari, and Renata~M. Wentzcovitch.
\newblock {{QUANTUM ESPRESSO}}: A modular and open-source software project for
  quantum simulations of materials.
\newblock {\em J. Phys.: Condens. Matter}, 21(39):395502, 2009.

\bibitem{vanderbiltSoftSelfconsistentPseudopotentials1990}
David Vanderbilt.
\newblock Soft self-consistent pseudopotentials in a generalized eigenvalue
  formalism.
\newblock {\em Phys. Rev. B}, 41(11):7892--7895, 1990.

\bibitem{perdewRestoringDensityGradientExpansion2008}
John~P. Perdew, Adrienn Ruzsinszky, G{\'a}bor~I. Csonka, Oleg~A. Vydrov,
  Gustavo~E. Scuseria, Lucian~A. Constantin, Xiaolan Zhou, and Kieron Burke.
\newblock Restoring the {{Density-Gradient Expansion}} for {{Exchange}} in
  {{Solids}} and {{Surfaces}}.
\newblock {\em Phys. Rev. Lett.}, 100(13):136406, 2008.

\bibitem{tingRefinedDFTMethod2023}
Yin-Ying Ting and Piotr~M. Kowalski.
\newblock Refined dft+$u$ method for computation of layered oxide cathode
  materials.
\newblock {\em Electrochimica Acta}, 443:141912, 2023.

\bibitem{bruggerOutputStatisticsQuantum2022}
Jonathan Brugger, Christian Seidel, Michael Streif, Filip~A. Wudarski,
  Christoph Dittel, and Andreas Buchleitner.
\newblock Output statistics of quantum annealers with disorder.
\newblock {\em Phys. Rev. A}, 105(4):042605, 2022.

\bibitem{benedettiEstimationEffectiveTemperatures2016}
Marcello Benedetti, John {Realpe-G{\'o}mez}, Rupak Biswas, and Alejandro
  {Perdomo-Ortiz}.
\newblock Estimation of effective temperatures in quantum annealers for
  sampling applications: {{A}} case study with possible applications in deep
  learning.
\newblock {\em Phys. Rev. A}, 94(2):022308, 2016.

\bibitem{benedettiQuantumAssistedLearningHardwareEmbedded2017}
Marcello Benedetti, John {Realpe-G{\'o}mez}, Rupak Biswas, and Alejandro
  {Perdomo-Ortiz}.
\newblock Quantum-{{Assisted Learning}} of {{Hardware-Embedded Probabilistic
  Graphical Models}}.
\newblock {\em Phys. Rev. X}, 7(4):041052, 2017.

\bibitem{saubanereIntuitiveEfficientMethod2014}
M.~Sauban{\`e}re, M.~Ben Yahia, S.~Leb{\`e}gue, and M.-L. Doublet.
\newblock An intuitive and efficient method for cell voltage prediction of
  lithium and sodium-ion batteries.
\newblock {\em Nat Commun}, 5(1):5559, 2014.

\end{thebibliography}

\end{document}